\documentstyle[aps,prb,amsfonts,amssymb,floats]{revtex}
\begin{document}
\draft
\wideabs{
\title{Screened-interaction expansion for the Hubbard model\protect\\and
determination of the quantum Monte Carlo Fermi surface}
\author{G\"{o}khan Esirgen\cite{address_GE,email_GE}
and Heinz-Bernd Sch\"{u}ttler}
\address{Center for Simulational Physics, Department of Physics and Astronomy,
University of Georgia, Athens, Georgia 30602-2451}
\author{Carsten Gr\"{o}ber and Hans Gerd Evertz\cite{address_HGE}}
\address{Institut f\"{u}r Theoretische Physik, Am Hubland, Universit\"{a}t
W\"{u}rzburg, D-97074 W\"{u}rzburg, Germany}
\date{\today}
\maketitle
\begin{abstract}
We develop a systematic self-consistent perturbative expansion for the self
energy of Hubbard-like models. The interaction lines in the Feynman diagrams
are dynamically screened by the charge fluctuations in the system. Although
the formal expansion is exact---assuming that the model under the study
is perturbative---only if diagrams to all orders are included, it is
shown that for large-on-site-Coulomb-repulsion-$U$ systems weak-coupling
expansions to a few orders may already converge.
In order to test the approximation
at intermediate-to-high temperatures, we use the exact charge-fluctuation
susceptibility from quantum Monte Carlo (QMC)
simulation studies as input, which
determines the exact screened interaction, and compare our results for the
self energy to the QMC results. We also make comparisons with
fluctuation-exchange (FLEX) approximation. We show that
the screened interaction for the large-$U$ system
can be vanishingly small at a certain intermediate
electron filling; and it is found that
our approximation for the imaginary part of
the one-particle self energy agrees well with the QMC results in
the low energy scales at this particular filling.
But, the usefulness of the approximation is
hindered by the fact that it has the incorrect filling dependence
when the filling deviates from this value. We also calculate the exact
QMC Fermi surfaces for the two-dimensional (2-D)
Hubbard model for several fillings. Our results near
half filling show extreme
violation of the concepts of the band theory;
in fact, instead of growing, Fermi surface vanishes when doped
toward the half-filled
Mott-Hubbard insulator. Sufficiently away from half filling,
noninteracting-like Fermi surfaces are recovered. These results combined
with the Luttinger theorem might show that diagrammatic expansions for
the nearly-half-filled Hubbard model are unlikely to be possible;
however, the nonperturbative part of the solution seems to be
less important as the filling gradually moves away from one half.
Results for the 2-D one-band Hubbard model for
several hole dopings are presented. Implications of this study
for the high-temperature superconductors are also discussed.
\end{abstract}
}

\section{INTRODUCTION}
\label{sec_int}

The basic Hamiltonian for the simplest description of interacting electrons
in a periodic potential of fixed lattice ions, the one-band Hubbard model,
was widely investigated after the discovery of high-temperature
superconductors.~\cite{hubbard}
In two dimensions, this model is widely accepted to have
an antiferromagnetic ground state at half filling of its tight-binding band,
although the temperature evolution and the relation with
the antiferromagnetism of its insulating electron spectrum
is poorly understood.~\cite{mis}
But, there is still ongoing debate about the nature
of this model, in fact, there are various mysteries, at band fillings
close but not equal to one half. The model may have a superconducting
ground state or be close to such an instability at these fillings;
therefore, understanding the Hubbard model seems to be crucial for an
understanding of the high-temperature superconductors.

There have been numerous approaches to the solution of the
two-dimensional (2-D) Hubbard model.
Exact diagonalization studies are limited to very small lattices and
mainly for this reason they are inconclusive.~\cite{ed}
Because of statistical errors,
exact quantum Monte Carlo (QMC)
studies at band fillings corresponding
to that of the high-temperature superconductors (about $15\%$ doped away from
one half) are limited to temperatures no less than about a quarter of the
electron-hopping energy of the model.~\cite{qmc}
At such moderately high temperatures
little information can be obtained about possible low-temperature
instabilities such as superconductivity. But it should be reminded that
QMC studies have led to a fairly good understanding of the insulating
antiferromagnetic behavior of the half-filled model which corresponds to the
undoped parent compounds of the high-temperature superconductors.

To overcome the lattice size or temperature limitations of the available
exact methods, approximate methods are needed.
An important class of such approximations use diagrammatic formalisms.
Since these approximations are weak coupling in nature, it is important
that the approximation sums up the physically important diagrams that
constitute the exact
infinite perturbative expansion. For example, self-consistent
approximations like fluctuation-exchange (FLEX) approximation, generalizes
the physically important Hartree-Fock approximation by adding
electron-hole and electron-electron pair scattering events to the bare
Coulomb interaction.~\cite{flex1}

In order to go beyond Hartree-Fock- or FLEX-type approximations, one has to
employ additional diagrams. A nice formal approach to this is to
renormalize, i.e., replace with a perturbative expansion,
the bare-interaction lines in the diagrams. But, this can cause
an enormous numerical-calculational challenge, because now, the interaction
lines, in principle, can depend on the momentum and frequency of the
incoming and outgoing electrons, whereas for the bare Coulomb interaction,
they don't.

In this article, we present a new self-consistent
diagrammatic expansion for the
electron self energy, in which the interaction lines are renormalized.
The systematic perturbative expansion sums up the skeleton diagrams which
exclude electron loops that represent a contribution
to the charge fluctuations.
Renormalization of the interaction lines is
therefore achieved by the screening of the bare Coulomb interaction by
the charge fluctuations. Although the approximation is exact, assuming that
the model under the study is perturbative, only if
the diagrams to all orders are included, it may converge rapidly if the
screened interaction is considerably weaker than the bare Coulomb interaction.
For the large-on-site-Coulomb-repulsion-$U$ 2-D Hubbard model,
we evaluate the {\em exact\/} screened interaction
from the QMC data, and show that this is indeed the case near, but sufficiently
doped away from, half filling.
Then, by using this exact screened interaction,
we evaluate the electron self energy up to the third order, and compare to
the QMC results. We find that the imaginary part of the self energy at low
energies almost
converges to the exact QMC results at the third order.

This article is organized as follows: In Section~\ref{sec_met} we formulate
our approximation. In Section~\ref{sec_self} we present and discuss
our results for the one-particle self energy
along with the exact QMC and approximate FLEX calculations.
We also calculate the exact QMC Fermi surfaces of the 2-D Hubbard model
and discuss them in Section~\ref{sec_fs}.
And finally, we summarize our main findings in Section~\ref{sec_sum}.

\section{METHOD}
\label{sec_met}

We will develop our diagrammatic expansion for the electron self energy
for the case of the 2-D one-band extended Hubbard model. The expansion is valid
in other dimensions as well and is easily generalizable to multiband
models. The Hamiltonian is
\begin{equation}
{\cal H}=\sum_{ ij}\left(-\sum_{\sigma}
t_{ij} c^{\dagger}_{i\sigma} c_{j\sigma}
+\frac{1}{2}\sum_{\sigma\sigma'}  V_{ij}n_{i\sigma} n_{j\sigma'}\right),
\label{eq1}
\end{equation}
where $c^{\dagger}_{i\sigma}$ creates an electron
with spin $\sigma$ at site ${\bf R}_{i}$
on an $N\!=\!L\!\times\!L$ square lattice with
periodic boundary conditions, and
$n_{i\sigma}=c^{\dagger}_{i\sigma}c_{i\sigma}$.
The lattice constant, $a$, is set to be $1$.
The Coulomb matrix element $V_{ij}\!\equiv\!V({\bf R}_{i}-{\bf R}_{j})$ 
comprises the Hubbard on-site repulsion 
$U\equiv V({\bf 0})$ and an extended part,
$V({\bf \Delta R})$, for non-zero ${\bf \Delta R}$.
For the calculations done in this article,
we include only a first-neighbor hybridization $t$
and the chemical potential $\mu$ in $t_{ij}$, and the on-site $U$
in $V({\bf \Delta R})$.
For the high-temperature-superconducting cuprates, it is estimated that
$t\sim0.3$--$0.5$~eV and $U/t\sim8$--$12$.\cite{HSSJ,ScFe}
We will use $U/t=8$ in all our calculations.

\begin{figure}[hbtp]
\begin{picture}(3.375,3.000)
\includegraphics{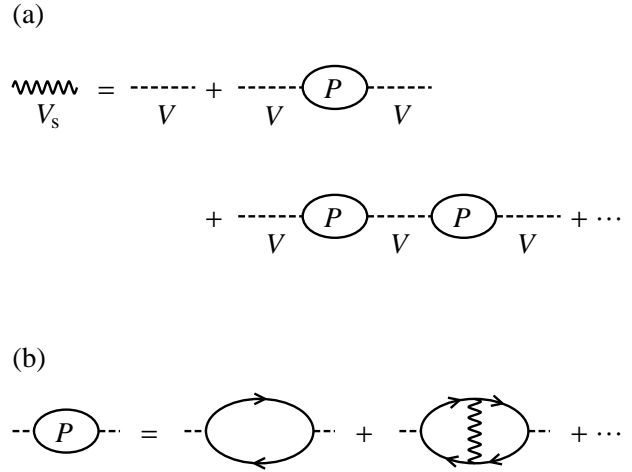}
\end{picture}
\caption{(a)~Exact screened interaction, $V_{\rm s}$\@. $V$ is the bare Coulomb
interaction. (b)~Exact polarization insertion, $P$\@.}
\label{fig_vsp}
\end{figure}

In Fig.~\ref{fig_vsp}(a), we show the generic form of the {\em exact\/}
screened interaction,
$V_{\rm s}$, expressed in terms of the bare Coulomb interaction, $V$, and
the {\em exact\/} polarization insertion, $P$\@. In momentum-frequency space,
\begin{equation}
V_{\rm s}(Q)=\frac{V(Q)}{1-P(Q)V(Q)},
\label{eq_vsp}
\end{equation}
where $Q=({\bf Q},i\Omega)$.
Fig.~\ref{fig_vsp}(b) shows the zeroth- and first-order diagram contributions
to $P$, expressed in terms of $V_{\rm s}$. This form of expression is useful
because it allows one
to write down a self-consistent approximation for $V_{\rm s}$ and $P$, given
a one-particle Green's function.

These two simply-related quantities are also similarly related to a physical
quantity easily measurable in  QMC studies, the charge susceptibility,
$\chi_{\rm c}$. The expression for this two-particle correlation function
as a thermodynamic expectation value is
\begin{eqnarray}
\chi_{\rm c}({\bf Q},i\Omega)=\frac{1}{N}
\sum_{ij}\int_{0}^{\beta}d\tau\, & & e^{i\Omega
\tau-i{\bf Q}.({\bf R}_i-{\bf R}_j)}\nonumber\\
 & \lefteqn{\times\langle\Delta n_{i}(\tau)\Delta n_{j}(0)\rangle,} &
\label{eq_xcexp}
\end{eqnarray}
where $n_{i}(\tau) \equiv \sum_{\sigma}
c^{\dagger}_{i\sigma}(\tau)c_{i\sigma}(\tau)$, and
$\Delta n_{i}(\tau) \equiv n_{i}(\tau)-\langle n_{i}(\tau) \rangle$.

\begin{figure}[hbtp]
\begin{picture}(3.375,0.750)
\includegraphics{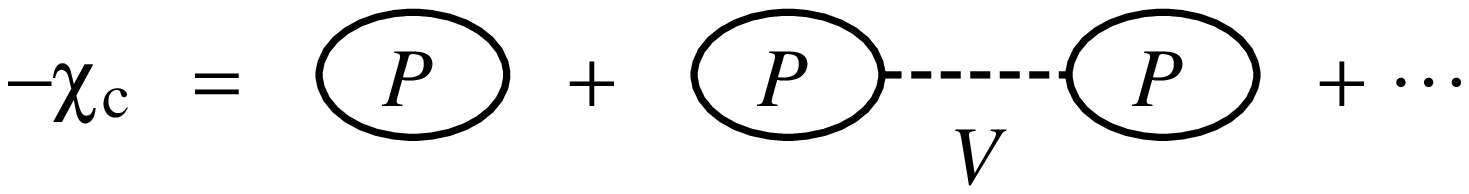}
\end{picture}
\caption{The charge susceptibility, $\chi_{\rm c}$, expanded in terms of
the polarization insertion, $P$\@. $V$ is the bare Coulomb interaction.}
\label{fig_xc}
\end{figure}

In principle, the definition of a polarization insertion, $P$,
is arbitrary, and so is of a screened interaction, $V_{\rm s}$, through
Eq.~(\ref{eq_vsp}). $P$ defined in this article through
the diagrammatic expansion in Fig.~\ref{fig_vsp}(b)
actually corresponds to the polarization insertion of the charge
susceptibility, $\chi_{\rm c}$. The exact $\chi_{\rm c}$ is expressible as a
geometric series in terms of the exact $P$, such that
\begin{equation}
-\chi_{\rm c}(Q)=\frac{P(Q)}{1-P(Q)V(Q)}
\label{eq_xcp}
\end{equation}
(see also Fig.~\ref{fig_xc}).
Combining Eqs.~(\ref{eq_vsp}) and~(\ref{eq_xcp}), we obtain an important
expression for the screened interaction, $V_{\rm s}$, in terms of the
{\em physical\/} charge susceptibility of the system, $\chi_{\rm c}$;
\begin{equation}
V_{\rm s}(Q)=V(Q)-V(Q)\chi_{\rm c}(Q)V(Q).
\label{eq_vsxc}
\end{equation}

\begin{figure}[hbtp]
\begin{picture}(3.375,6.600)
\includegraphics{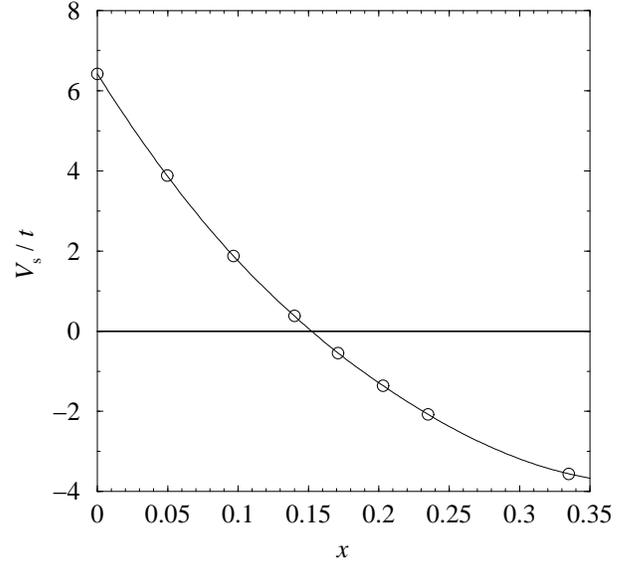}
\end{picture}
\caption{$V_{\rm s}({\bf \Delta R}={\bf 0},i\Omega=0)$ as a
function of the hole doping, $x\equiv 1-\langle n \rangle$,
obtained from QMC
for the 2-D Hubbard model. Parameters are $U/t=8$; $T/t=1/3$, where $T$
is the temperature; and the lattice size is $8\times8$.}
\label{fig_vsqmc}
\end{figure}

At this point, we are ready to calculate the exact $V_{\rm s}$ by using
the exact $\chi_{\rm c}$ obtained by QMC. Using $V(Q)=U$ for the Hubbard
model, we have,
\begin{equation}
V_{\rm s}(Q)=U-U^{2}\chi_{\rm c}(Q).
\label{eq_vshub}
\end{equation}
One can also Fourier transform $V_{\rm s}(Q)$ to the real space, by using,
\begin{equation}
V_{\rm s}({\bf \Delta R},i\Omega)=\frac{1}{N}\sum_{\bf Q}
e^{i{\bf Q}.{\bf \Delta R}}V_{\rm s}({\bf Q},i\Omega).
\label{eq_vsdr}
\end{equation}
Again, for the Hubbard model,
\begin{equation}
V_{\rm s}({\bf \Delta R},i\Omega)=
U\delta({\bf \Delta R})-U^{2}\chi_{\rm c}({\bf \Delta R},i\Omega),
\label{eq_vshubdr}
\end{equation}
where $\delta$ is the Kronecker delta. In Fig.~\ref{fig_vsqmc}, we
plot the on-site and time-independent component of the exact screened
interaction obtained from QMC, $V_{\rm s}({\bf \Delta R}={\bf 0},i\Omega=0)$,
for different hole dopings, $x\equiv 1-\langle n \rangle$, measured
with respect to half filling.
Although the bare (unscreened) Coulomb interaction, $U$, is equal to
$8t$, strikingly, the corresponding
component of the exact screened interaction is substantially weaker for
intermediate hole dopings, $x$, actually vanishing and changing
sign, i.e., becoming attractive, at $x\sim15\%$. We will refer to this effect
as ``overscreening,''~\cite{sgeh} and explain below why it happens.

From Eq.~(\ref{eq_vshubdr}), because of the $U^2$ term, one can suspect that
for sufficiently large $U$, this term will win and
$V_{\rm s}({\bf \Delta R}={\bf 0},i\Omega=0)$ will become negative
(attractive), since $\chi_{\rm c}({\bf \Delta R}={\bf 0},i\Omega=0)$ is
always positive and approaches a nonzero $U$-independent limit
[of ${\cal O}(x/t)$ by a simple $U=\infty$ scaling argument] for $U\gg t$.
On the other hand, at half filling, charge fluctuations are suppressed;
so that $\chi_{\rm c}({\bf \Delta R},i\Omega=0)\sim {\cal O}(t^2 /U^3)$ and
$V_{\rm s}({\bf \Delta R}={\bf 0},i\Omega=0)\simeq U$, for $U \gg t$.
Hence, presence of a sufficiently large $U$ {\em and\/} finite hole
doping off half filling, $x$, is necessary for
overscreening. Also, note that overscreening is not restricted to the on-site
component of the Coulomb interaction; the nearest-neighbor
components, and so forth,
can also become attractive because of the overscreening
effect.~\cite{sgeh}

At this point, we are ready to write down a controlled perturbative expansion
for the one-particle self energy. Since the exact
screened interaction, $V_{\rm s}$, was shown to be substantially weaker than
the bare interaction, the Hubbard $U$, one can expect that an expansion
in terms of $V_{\rm s}$ should converge much more rapidly than a
brute-force expansion in terms of $U$. As a matter of fact, it is not difficult
to pick up the diagrams which contribute to this expansion. Keeping in mind
that our approximation for the self energy will be self consistent in the
sense that all one-particle-Green's-function lines will correspond
to the full Green's function, we should allow only the skeleton diagrams
in which none of these lines have any self-energy diagrams explicitly attached
to them. As a second step, bearing in mind the fact that the interaction
lines in our diagrams, which correspond to $V_{\rm s}$, already include
the polarization
diagrams in them [Fig.~\ref{fig_vsp}(a)], we omit any self-energy diagram
in which an interaction line
has a polarization diagram [see Fig.~\ref{fig_vsp}(b)] inserted.
These are basically the only rules needed in the diagram selection.
If all the diagrams to infinite order are included,
an {\em exact\/} perturbation
series will result, like in any other rigorous perturbative expansion.
Note that the interaction lines in our diagrams,
or $V_{\rm s}$, represent the bare Coulomb interaction, $U$, screened by
the charge fluctuations, $\chi_{\rm c}$~[Eq.~(\ref{eq_vsxc})]. Therefore,
we have derived a diagrammatic expansion in terms of the charge fluctuations,
which is exact if carried out to all orders. Moreover, because of the
overscreening effect discovered by the analysis of the interaction
lines ($V_{\rm s}$) obtained by QMC exactly,
a weak-coupling expansion to a first
few orders is expected to converge in a rapid, controlled fashion.

\begin{figure}[hbtp]
\begin{picture}(3.375,4.400)
\includegraphics{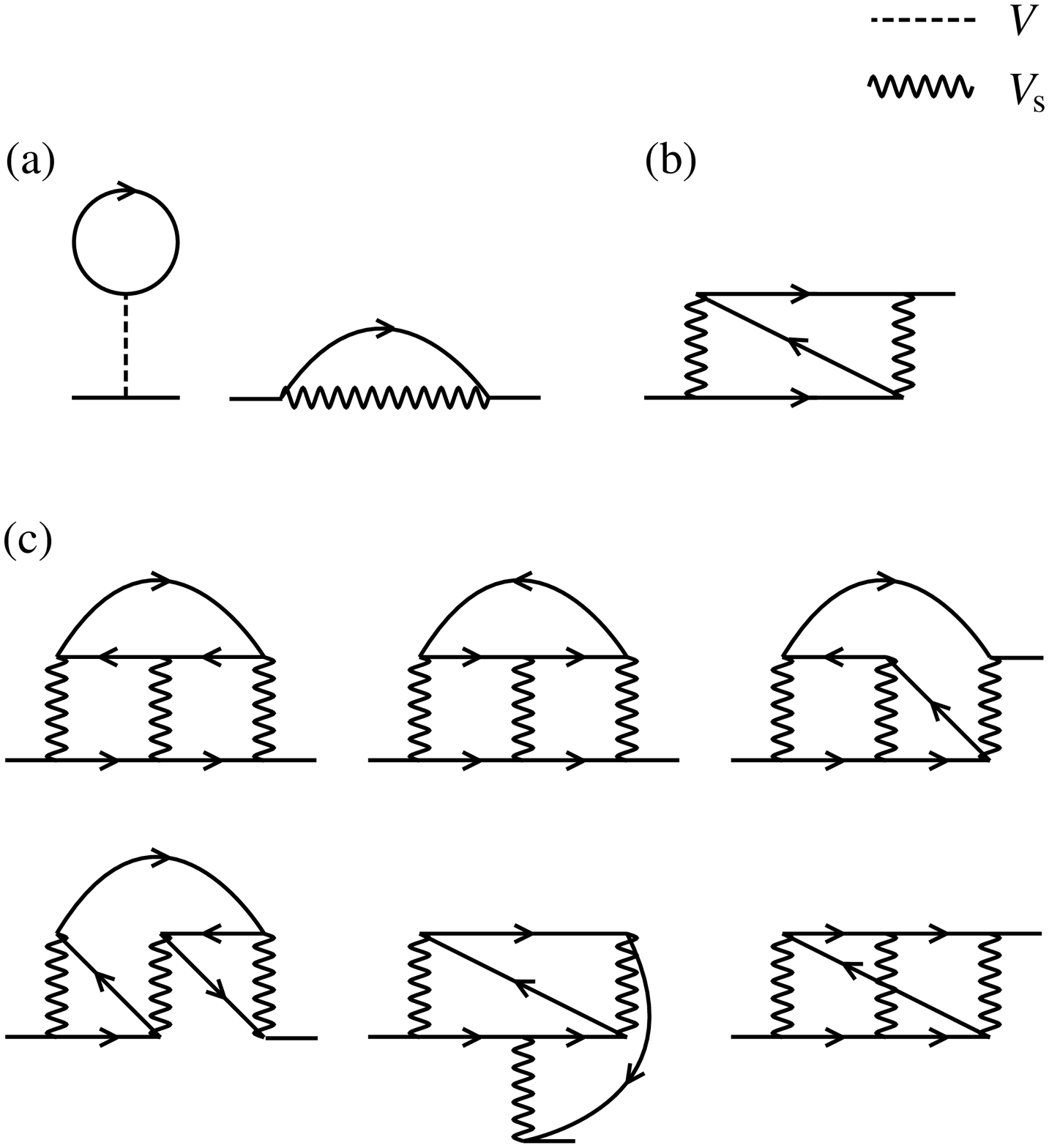}
\end{picture}
\caption{All the contributing self energy diagrams in
the first~(a), second~(b), and third order~(c) of the screened-interaction
($V_{\rm s}$) expansion. The weak-coupling expansion would become exact
(assuming the model under study is perturbative) when
carried out to all orders.}
\label{fig_sig}
\end{figure}

In Fig.~\ref{fig_sig}, we show all the self-energy diagrams up to the third
order. The Hartree diagram in Fig.~\ref{fig_sig}(a) is written separately
in terms of the bare interaction ($U$ in this case) to prevent double
counting; for the one-band Hubbard model, it is a trivial constant which
is equal to $U\langle n \rangle$.
The other first order diagram in Fig.~\ref{fig_sig}(a) is like
the Fock diagram with the bare interaction line replaced by the
screened interaction. This diagram combines the bare-interaction
Fock diagram (which is again equal to a trivial constant,
$-U\langle n \rangle / 2$, in this case) with
the charge-fluctuation self energy diagram found in FLEX-like approximations;
because, the screened interaction is the sum of the bare interaction
and the charge-fluctuation propagator [see, e.g., Eq.~(\ref{eq_vsxc})
or~(\ref{eq_vshub})]. In the second order, there is only one diagram
contributing to the self energy [Fig.~\ref{fig_sig}(b)]; and in the third
order, the number of all the contributing diagrams is six
[Fig.~\ref{fig_sig}(c)].

\begin{figure}[hbtp]
\begin{picture}(3.375,3.500)
\includegraphics{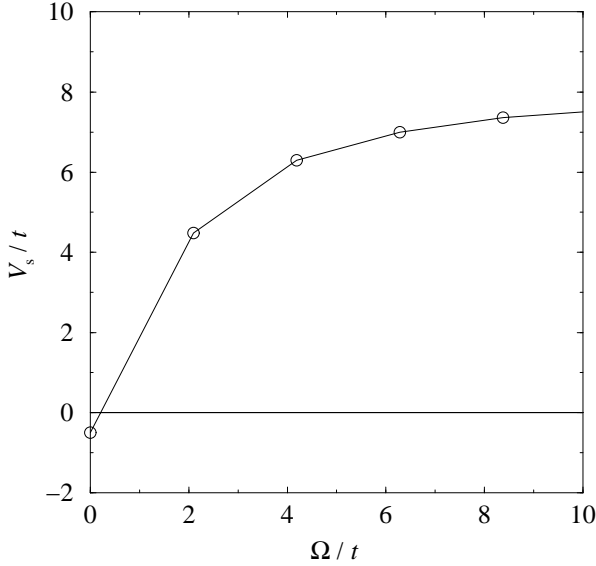}
\end{picture}
\caption{The exact screened interaction, $V_{\rm s}({\bf Q}, i\Omega)$,
calculated by QMC, shown as a function of the frequency, $\Omega$.
The momentum point is ${\bf Q}=(\pi/2,0)$ on an $8\times8$ lattice.
Parameters are $U/t=8$; $T/t=1/3$, where $T$
is the temperature; and $x=14.0\%$, where $x \equiv 1-\langle n \rangle$.}
\label{fig_vsw}
\end{figure}

At this point, we would like to
discuss the fact that, the screened-interaction expansion
developed here is a low-energy approximation. In Fig.~\ref{fig_vsw}, we
plot the exact screened interaction, $V_{\rm s}({\bf Q}, i\Omega)$,
which was obtained by QMC, as a function of the frequency, $\Omega$.
For the chosen ${\bf Q}$ point, $V_{\rm s}$ is actually attractive and
small at zero frequency, and vanishes at a slightly higher frequency.
But as we approach the first Matsubara frequency, $V_{\rm s}$ rapidly
grows, and it is already more than half of the bare, unscreened Coulomb
repulsion. As the frequency increases further, $V_{\rm s}$ finally approaches
to the bare repulsive interaction, $U=8t$. It is now clear that the
high-energy part of the $V_{\rm s}$ is not weak, and our approximation
is limited to the low-energy scales. Therefore, for the temperature
studied here ($T=t/3$), the calculated self energy is expected
to be accurate only for the first Matsubara frequency, $\pi T\approx t$,
a small-enough fraction of the width of the tight-binding band,
which governs the energy scale for $V_{\rm s}$.

Before finishing this section, we would like to discuss a peculiar property
of the irreducible polarization insertion and the associated three-point
vertex function. As we discussed earlier, the exact screened interaction,
$V_{\rm s}$, as obtained from QMC, vanishes at particular
$({\bf \Delta R},i\Omega)$ points, and it will vanish at particular
$({\bf Q},i\Omega)$ points likewise; this was the motivation behind
developing a weak-coupling expansion in terms of $V_{\rm s}(Q)$.
But from Eq.~(\ref{eq_vsp}), $V_{\rm s}(Q)=0$ immediately implies
that $P(Q)=\infty$ at the same $Q=({\bf Q},i\Omega)$ point.
This divergence in the polarization insertion of the charge susceptibility
does not imply a divergence in the charge susceptibility itself, which is
the actual physical quantity. In fact, from Eq.~(\ref{eq_xcp}),
$P(Q)=\infty$ simply implies that $\chi_{\rm c}(Q)=1/V(Q)$.
Although this divergence is not a worry from a physics point of view,
it imposes technical difficulties on the expansion of $P(Q)$
[see Fig.~\ref{fig_vsp}(b)] along these singularities;
to overcome this difficulty in an actual calculation of $P(Q)$,
it would be more appropriate to expand its inverse, $1/P(Q)$,
rather than to expand $P(Q)$ itself directly.

\begin{figure}[hbtp]
\begin{picture}(3.375,3.250)
\includegraphics{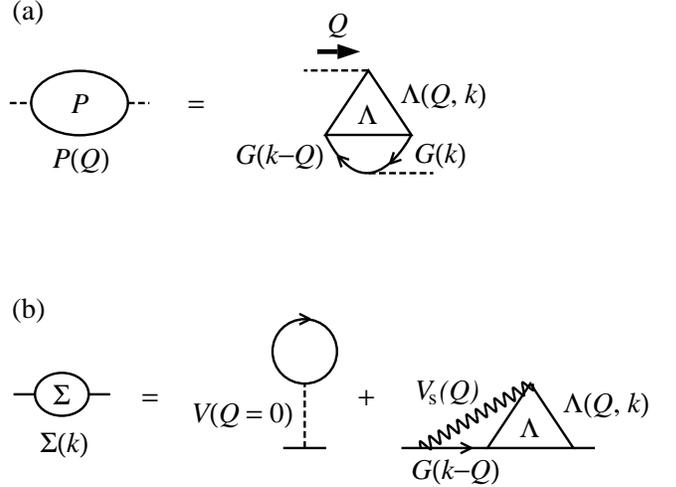}
\end{picture}
\caption{The polarization insertion, $P(Q)$~(a),
and the one-particle self energy, $\Sigma(k)$~(b),
in terms of the three-point irreducible vertex, $\Lambda(Q,k)$.
$G(k)$ is the one-particle Green's function, $Q=({\bf Q},i\Omega)$,
and $k=({\bf k},i\omega)$. $V$ and $V_{\rm s}$ are the bare and the screened
interaction respectively.}
\label{fig_lambda}
\end{figure}

Both the polarization insertion and the self energy can be expressed
in terms of a three-point irreducible vertex function,
$\Lambda(Q,k)$, by the exact relations (see Fig.~\ref{fig_lambda}),
\begin{equation}
P(Q)=\frac{2T}{N}\sum_{k}G(k)G(k-Q)\Lambda(Q,k),
\label{eq_lambdap}
\end{equation}
and
\begin{equation}
\Sigma(k)=V(Q=0)\langle n\rangle
-\frac{T}{N}\sum_{Q}\Lambda(Q,k)V_{\rm s}(Q)G(k-Q),
\label{eq_lambdasig}
\end{equation}
where $G(k)$ is the one-particle Green's function and $k=({\bf k},i\omega)$.
Eq.~(\ref{eq_lambdap}) immediately implies that for a given $Q$,
if $P(Q)=\infty$, then $\Lambda(Q,k)=\infty$, because $G(k)$ is always finite.
Even though $\Lambda(Q,k)$ in Eq.~(\ref{eq_lambdasig}) may diverge, the
one-particle self energy, $\Sigma(k)$, which is the actual physical quantity,
remains finite; because, if $\Lambda(Q,k)$ diverges
at a given $Q$, then $V_{\rm s}(Q)$
will vanish, therefore canceling the divergence. But, there is still a
concern that a weak-coupling expansion for $\Lambda$
in Fig.~\ref{fig_lambda}(b), which generates the diagrams
in Fig.~\ref{fig_sig}, may not be sufficient because of this
diverging behavior. We will put our theory to numerical testing in the
next section.

\section{ONE-PARTICLE SELF ENERGY}
\label{sec_self}

We perform our numerical calculations for the self energy on
periodic discreet lattices
of sizes $4\times4$ and $8\times8$, so that they can be compared with the QMC
results. We use the Matsubara-frequency representation, and employ a fermion
frequency cutoff of $\sim35t$. All the QMC calculations were done at a
temperature, $T=t/3$, and with the cutoff employed here, this corresponds to
16 positive Matsubara frequencies. Note that QMC calculations are performed
in the imaginary time, and a time discretization is made for numerical
purposes. The number of time slices in $[0,\beta=1/T]$ used in the
QMC calculations was 80, which is equivalent to using 40 positive
Matsubara frequencies---even larger a cutoff than our diagrammatic
calculations. We chose a smaller cutoff for our diagrammatic calculations
to avoid the high-frequency errors in the $V_{\rm s}$ obtained from QMC,
which goes directly into the calculations. We vary the
chemical potential, $\mu$, in order to match the filling, $\langle n\rangle$,
with that of the QMC. In order to calculate $\langle n\rangle$, the
one-particle Green's function, $G({\bf k},i\omega)$,
has to be summed over the Matsubara frequencies,
and to prevent cutoff effects in this relatively-slow-converging sum,
we effectively extend the sum to infinity by using the standard---replace
the high-frequency part of the sum with the result for the non-interacting
case---trick.

\begin{figure}[hbtp]
\begin{picture}(3.375,6.300)
\includegraphics{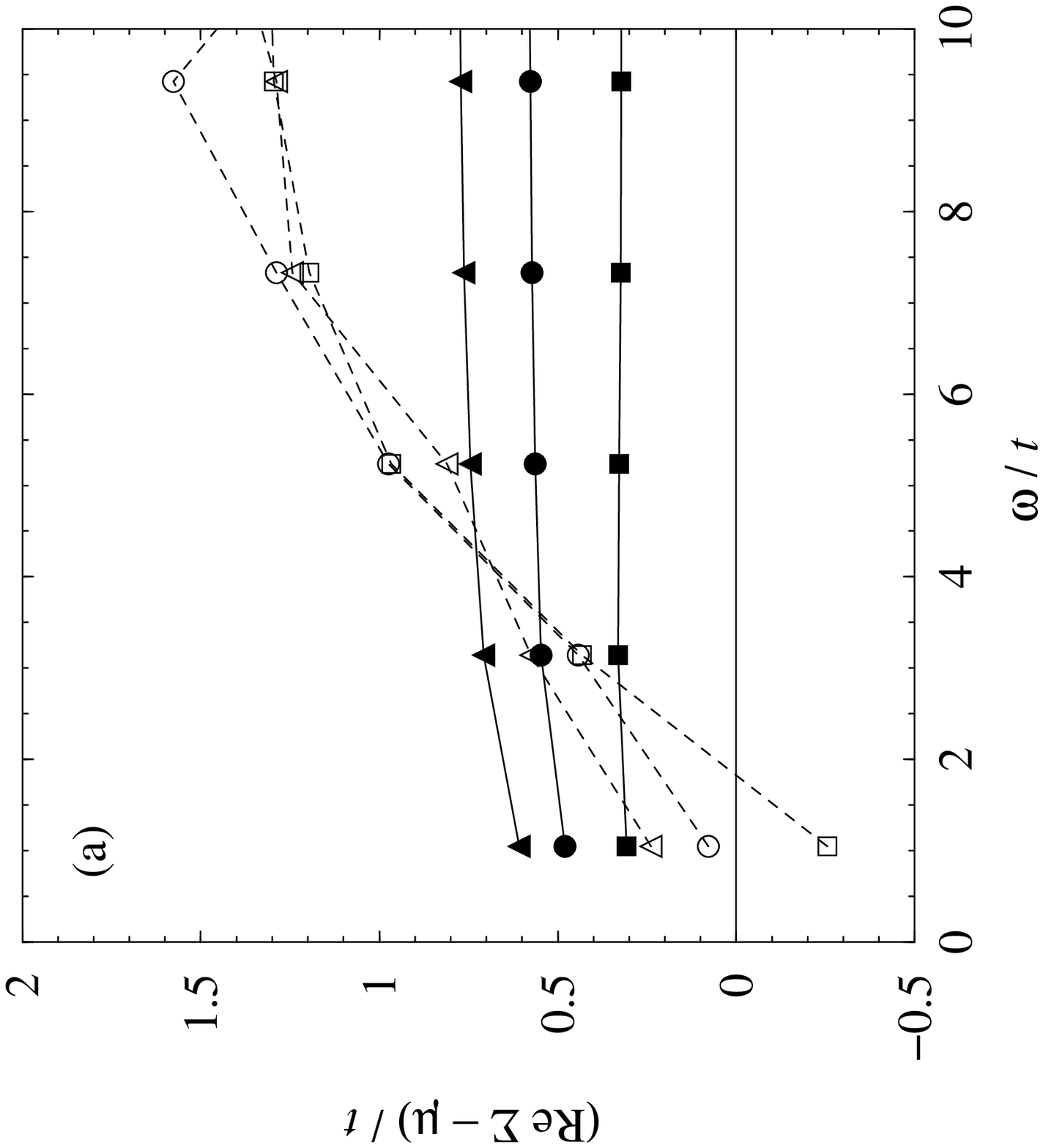}
\includegraphics{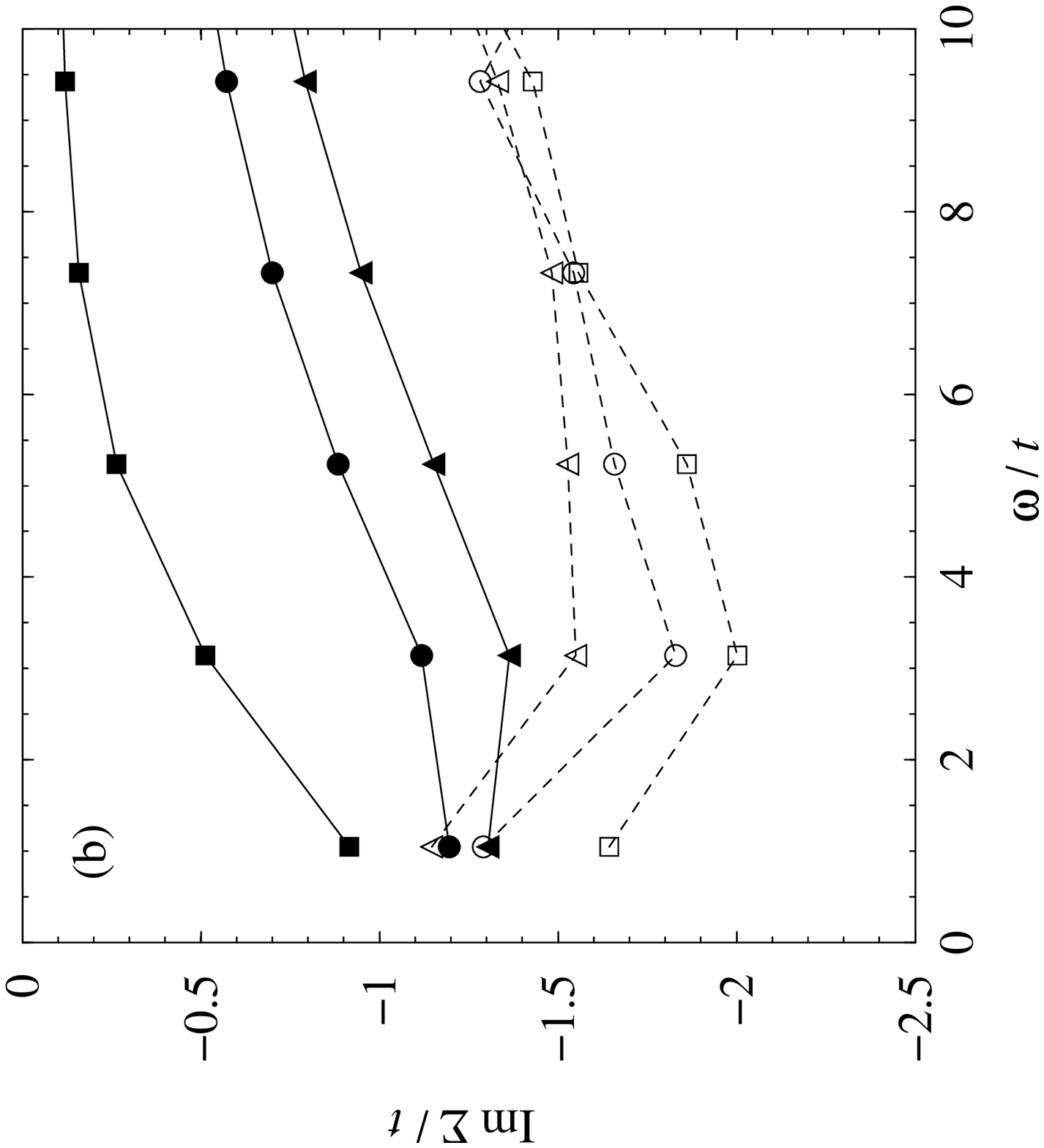}
\end{picture}
\caption{The one-particle self energy as a function of the Matsubara
frequency, shown for the screened-interaction expansion (solid lines
and filled symbols)
along with exact (except for statistical errors) QMC results
(dashed lines with empty symbols).
The ${\bf k}$ point is $(\pi,0)$.
Hole dopings,
$x\equiv 1-\langle n \rangle$,
are $10.0$ (squares), $14.2$ (circles), and $17.5\%$ (triangles).
Other parameters are $U/t=8$ and $T/t=1/3$. Lattice sizes are
$4\times4$ for the screened-interaction expansion and $8\times8$
for QMC. Hole dopings for both
the expansion and the QMC
correspond to averages on $4\times 4$ ${\bf k}$-space
meshes. Actual QMC hole dopings averaged on $8\times 8$ ${\bf k}$-space
meshes are $9.7$, $14.0$, and $17.1\%$.
The real~(a) and the imaginary~(b) part.
[We use $8\times 8$ QMC results in these comparisons because QMC calculations
were performed only on $8\times 8$ lattices for most hole dopings; but
the difference between $4\times 4$ and $8\times 8$ QMC results are well
within the error bars (see Fig.~\protect\ref{fig_sig_4X4} and
the associated discussion at the end of this section).]}
\label{fig_sig_scr}
\end{figure}

\begin{figure}[hbtp]
\begin{picture}(3.375,6.300)
\includegraphics{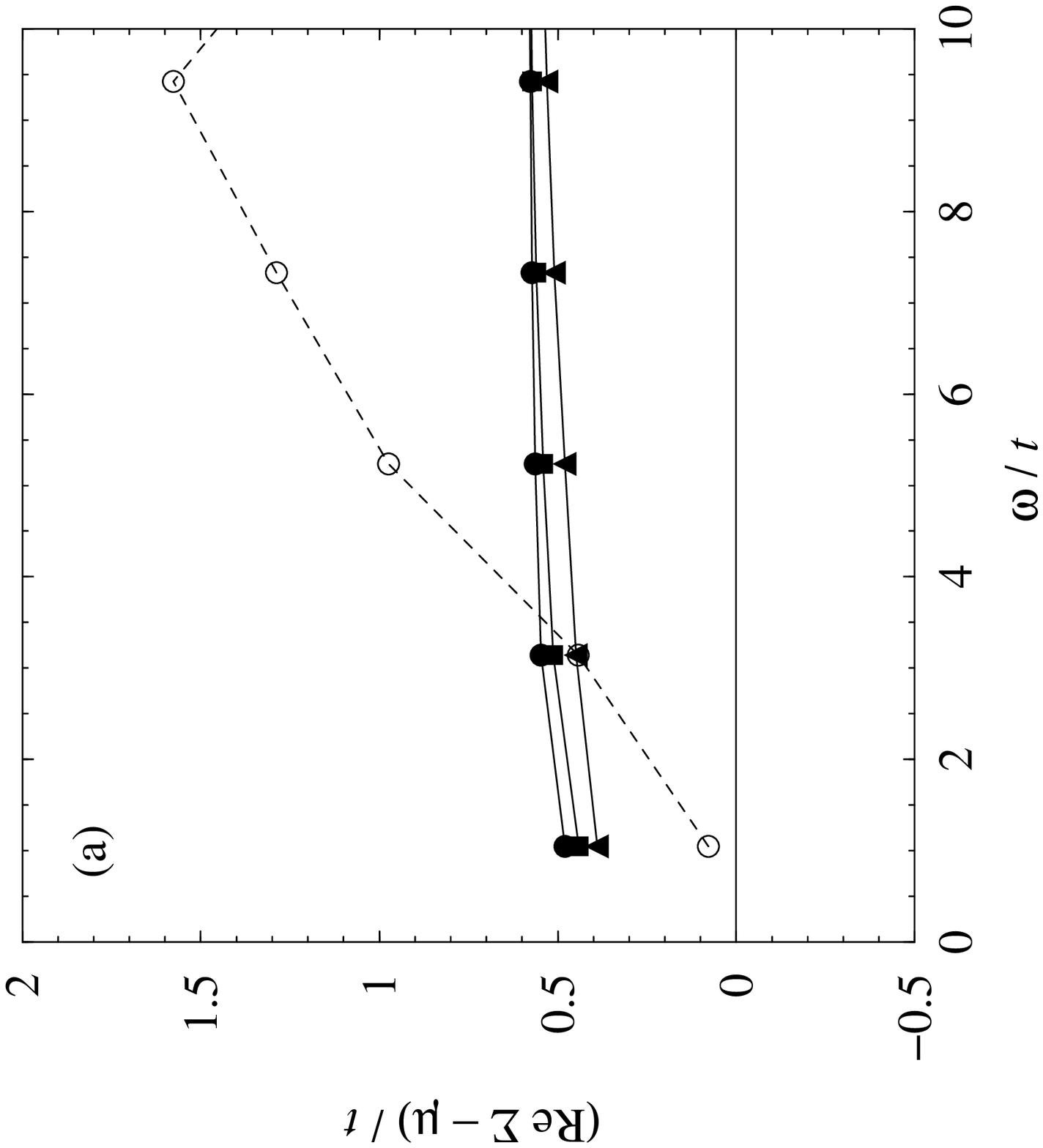}
\includegraphics{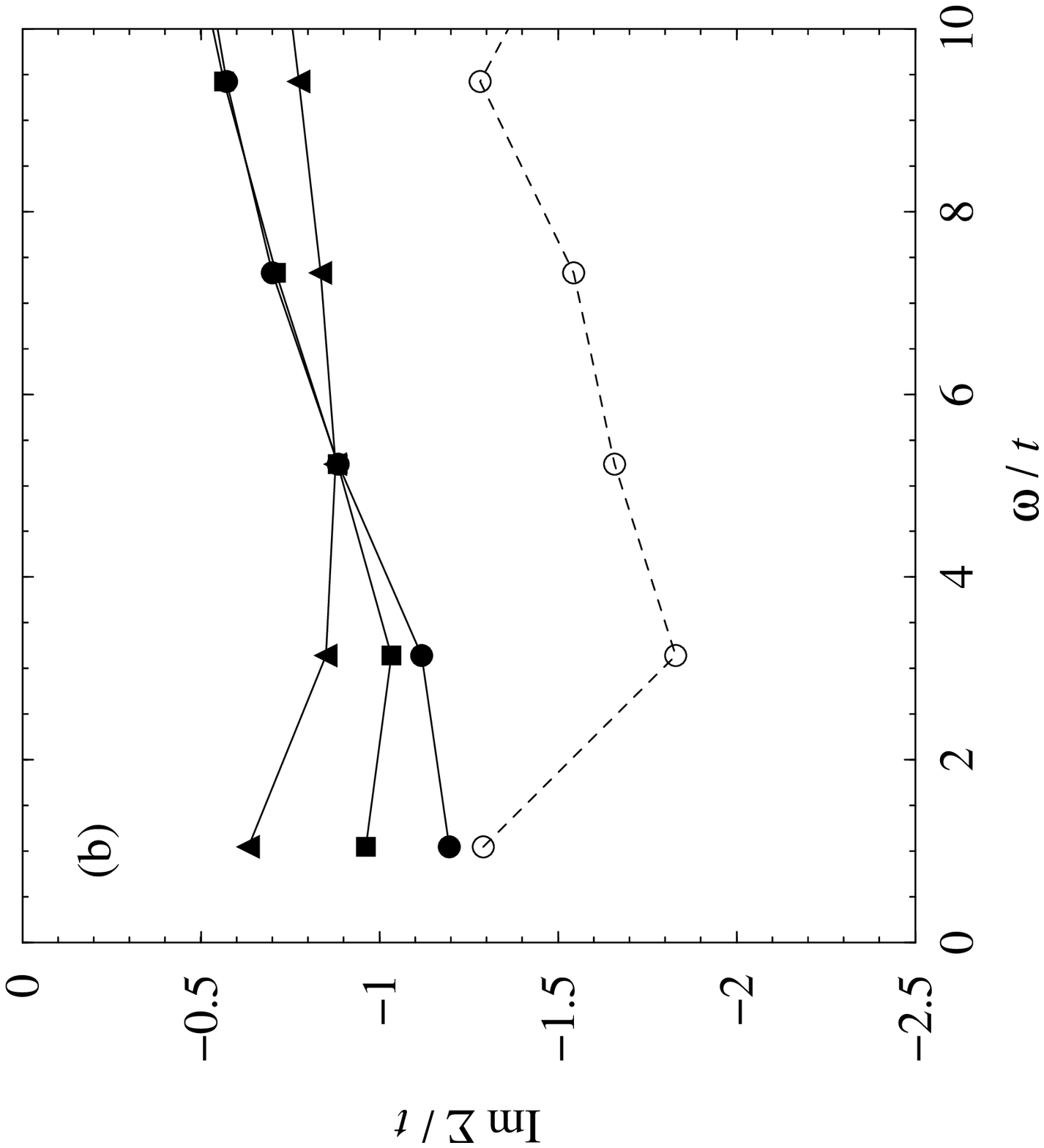}
\end{picture}
\caption{The one-particle self energy as a function of the Matsubara
frequency, shown for the 1st- (triangles), 2nd- (squares),
and 3rd-order (circles)
screened-interaction expansion (solid lines
and filled symbols)
along with exact (except for statistical errors) QMC results
(dashed lines and empty circles).
The ${\bf k}$ point is $(\pi,0)$.
Hole doping, $x\equiv 1-\langle n \rangle$,
is $14.2\%$.
All parameters are the same as in Fig.~\protect\ref{fig_sig_scr}.
The real~(a) and the imaginary~(b) part.}
\label{fig_sig_123}
\end{figure}

\begin{figure}[hbtp]
\begin{picture}(3.375,6.300)
\includegraphics{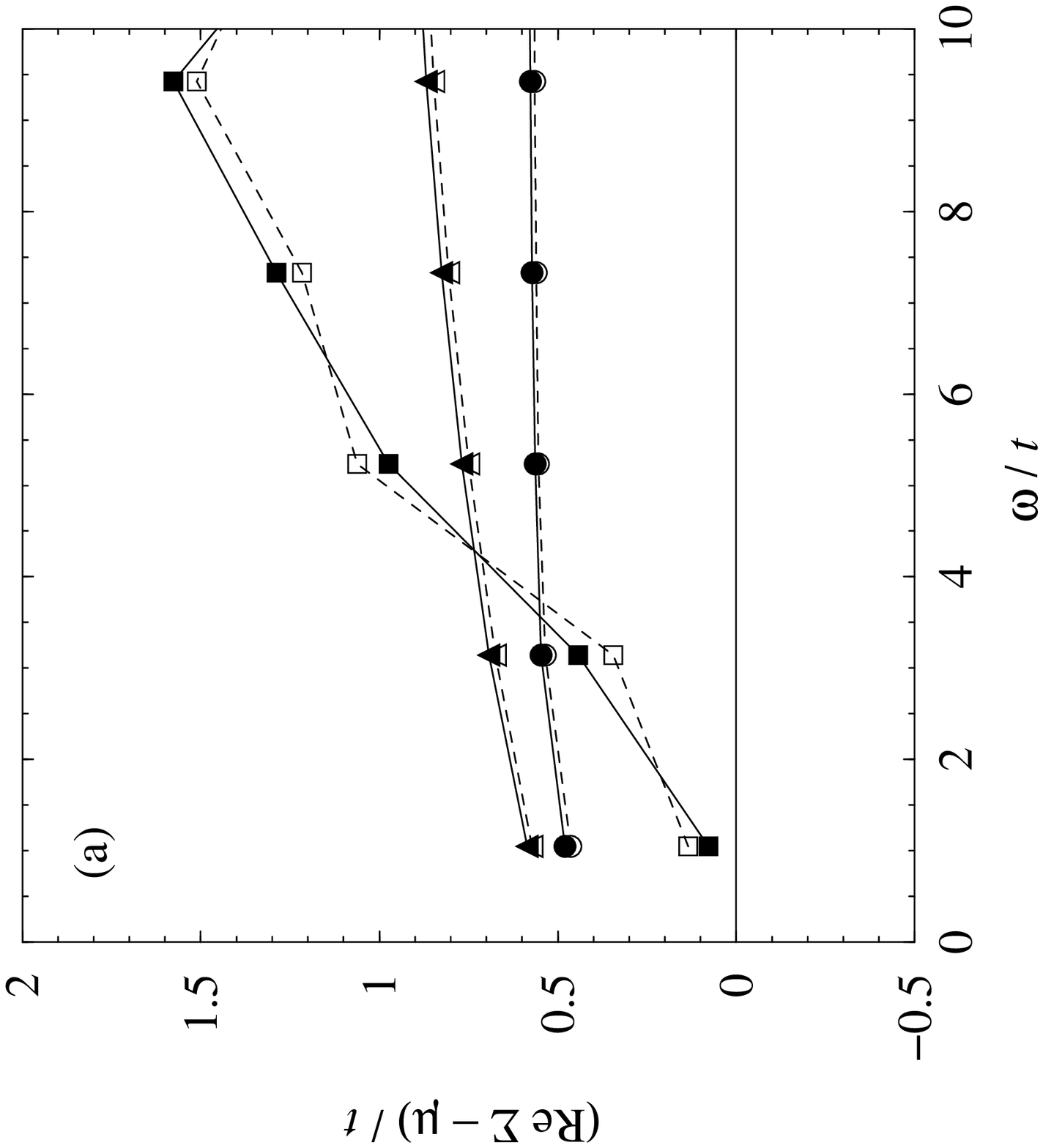}
\includegraphics{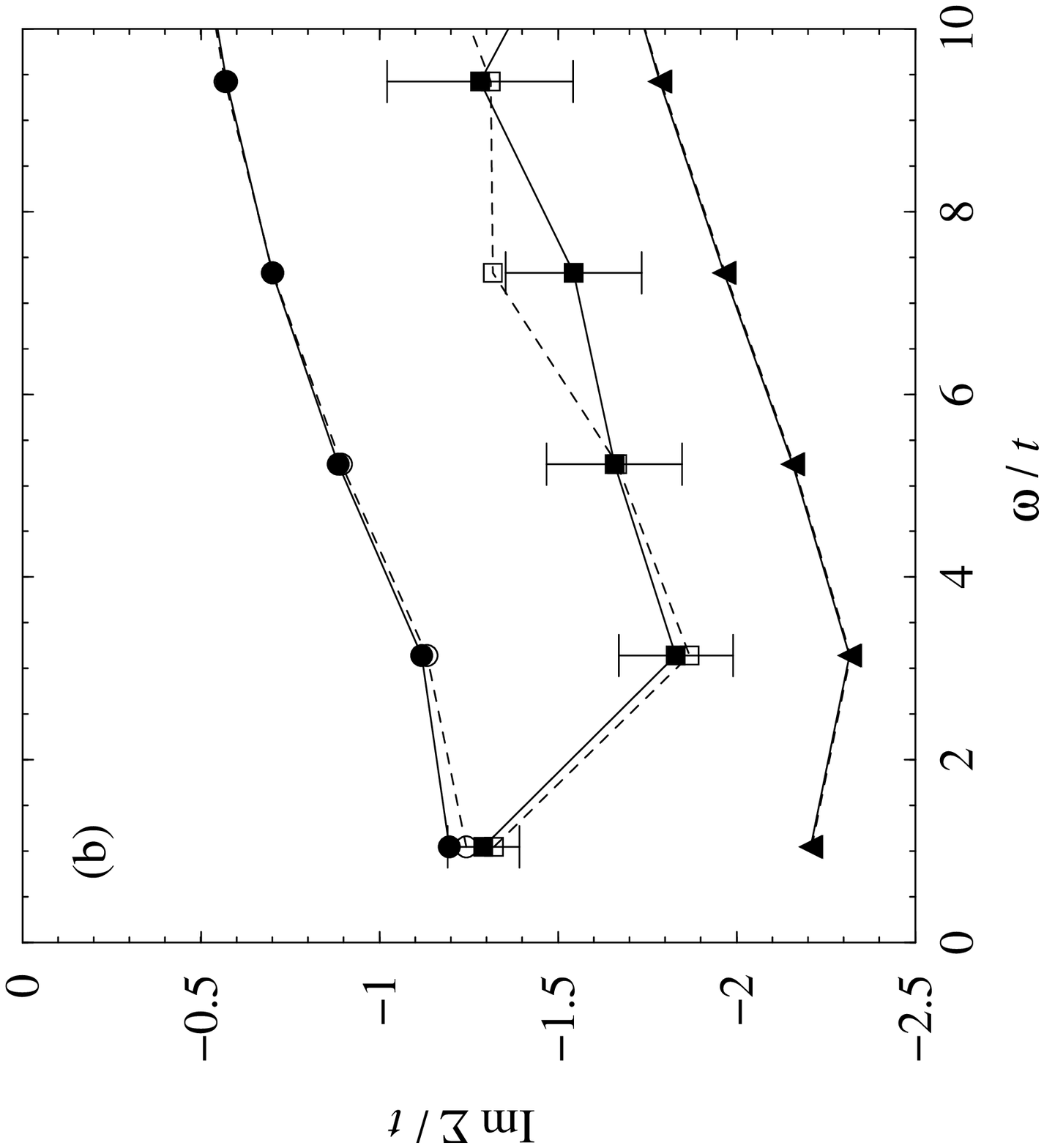}
\end{picture}
\caption{The one-particle self energy plotted as a function of the Matsubara
frequency for the purpose of evaluating lattice-size effects.
Results for QMC (squares),
screened-interaction expansion (circles), and FLEX approximation
(triangles) are shown.
The ${\bf k}$ point is $(\pi,0)$.
Solid lines and filled symbols represent $8 \times 8$ lattices
for QMC and $4 \times 4$ otherwise. Dashed lines and empty symbols
represent $4 \times 4$ lattices for all. Parameters are $U/t=8$ and $T/t=1/3$.
Hole dopings,
$x\equiv 1-\langle n \rangle$, averaged on $4\times 4$ ${\bf k}$-space meshes
are $14.2\%$ (solid lines and filled symbols) and $13.8\%$
(dashed lines and empty symbols). Actual, $8\times 8$-mesh averaged
hole doping for the $8\times 8$ QMC lattice is
$14.0\%$ (solid line with filled squares). Note that the
$4\times 4$-mesh averaged hole-doping
values for all data sets match each other
among the solid and the dashed lines.
The real~(a) and the imaginary~(b) part.
Also shown are the QMC error bars for the ${\rm Im\:}\Sigma$
on the $8\times 8$ lattice.}
\label{fig_sig_4X4}
\end{figure}

In Fig.~\ref{fig_sig_scr}, we plot the real and imaginary parts
of the self energy
as a function of Matsubara frequencies along with the results
from QMC calculations.
We chose a particular momentum point close to the
Fermi surface, ${\bf k}=(\pi,0)$,
but the ${\bf k}$ dependence of both the QMC and the
screened-interaction-expansion calculations near the Fermi surface is small.
QMC error bars are omitted
in all but one plot in this article
for clarity, but the estimated QMC error bar
for the real and imaginary parts of the self energy at the
first frequency is about $\pm0.1t$.
The corresponding error bar
for the QMC Green's function is about $\pm0.02t^{-1}$.
Also, the error bars for the self energy grow with the frequency,
whereas, they are roughly constant for the Green's function.
Note that, the temperature in these calculations is $T=t/3$,
and the second Matsubara Frequency, $5\pi T$, is already
about $5.2 t$, which is a substantial fraction of
the (non-interacting) bandwidth of our model, $8t$.
For the reasons we explained in the previous section, our approximation
is a low-energy approximation, which is unable to capture
this regime. But,
the first Matsubara frequency, $\pi T \approx 1.0t$,
is low enough compared to the bandwidth, and remarkably,
at the $14\%$ hole doping, where the screened interaction
almost vanishes (see Fig.~\ref{fig_vsqmc}),
our approximation
for the imaginary part of the self energy
almost converges to the QMC result within the error bar
[Fig.~\ref{fig_sig_scr}(b)].
This shows that our expansion may indeed be converging vary rapidly
at the hole doping at which the on-site component of the
screened interaction at zero frequency,
$V_{\rm s}({\bf \Delta R}={\bf 0},i\Omega=0)$, vanishes.
But, when the hole doping is varied away from this point,
${\rm Im\:}\Sigma$ varies in opposite direction with respect to
the exact QMC results. Naturally, having the correct doping
dependence is one of the most desired aspect of any approximation,
which, therefore, substantially limits the usefulness of this
expansion. It is quite possible that the reason for this
failure may be intimately related to the vertex divergence
problem discussed in the previous section.
In Fig.~\ref{fig_sig_123}, we show the evolution of the self energy
with the order of the diagrams included. The imaginary part, in fact,
seems to be converging to the QMC result at the lowest frequency
with the increasing order, and it is about the same as the QMC result
within the QMC error bar (see Fig.~\ref{fig_sig_4X4}) at the third order.
But the real part seems to be going in the wrong direction with
the increasing order.

The real part of the self energy
which was calculated by the screened-interaction expansion
actually does not do very well against
the exact QMC results [Fig.~\ref{fig_sig_scr}(a)] at all,
but, as we will analyze below, this is not a failure of the
expansion itself. In fact,
this is merely the quantity which determines the location of the
Fermi surface. Luttinger theorem\cite{luttinger}
states that, for any diagrammatic
expansion, the total volume (area in 2-D)
enclosed by the Fermi surface is equal to the
total filling, $\langle n \rangle$;
therefore our approximation
will obey this statement.
But then, this simply implies that the exact QMC results may not obey
the Luttinger theorem for the volume of the Fermi surface.
To investigate this further, we calculated the exact QMC Fermi surfaces
for various dopings, which will be discussed later.

\begin{figure}[hbtp]
\begin{picture}(3.375,6.300)
\includegraphics{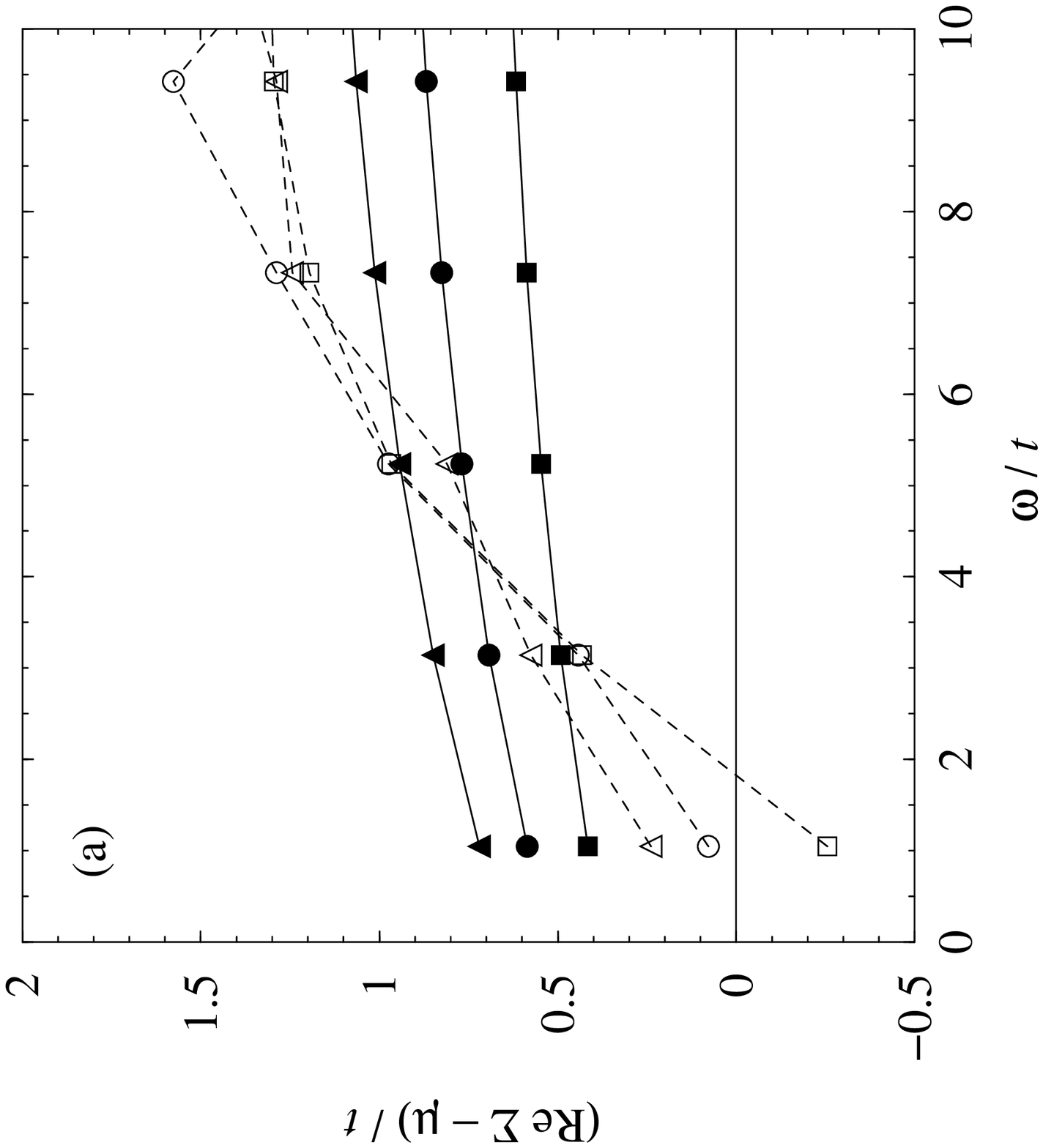}
\includegraphics{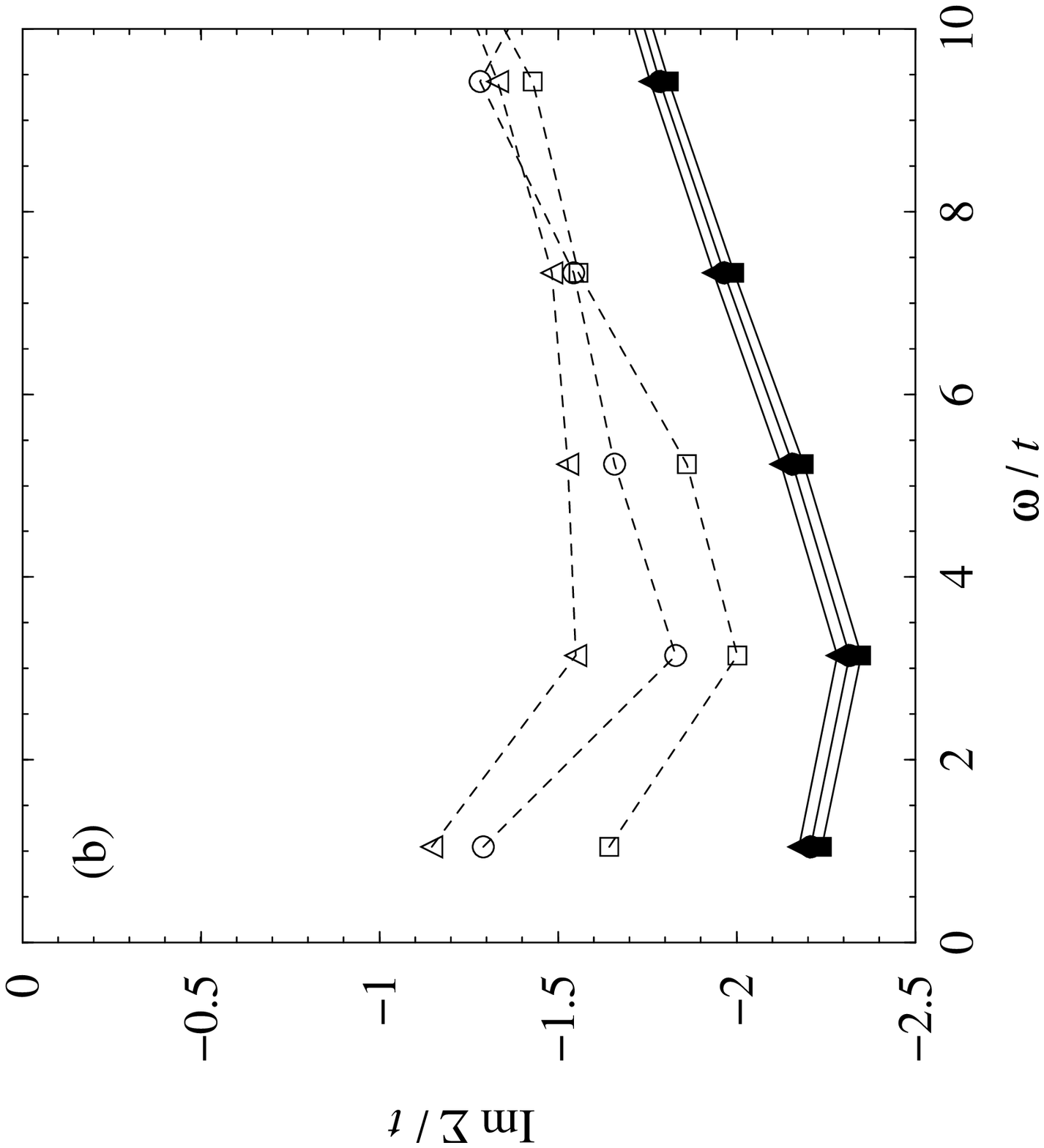}
\end{picture}
\caption{The one-particle self energy as a function of the Matsubara
frequency, shown for FLEX approximation (solid lines
and filled symbols)
along with exact (except for statistical errors) QMC results
(dashed lines with empty symbols).
The ${\bf k}$ point is $(\pi,0)$.
Hole dopings,
$x\equiv 1-\langle n \rangle$,
are $10.0$ (squares), $14.2$ (circles), and $17.5\%$ (triangles).
Other parameters are $U/t=8$ and $T/t=1/3$. Lattice sizes are
$4\times4$ for FLEX and $8\times8$
for QMC. Hole dopings for both the
FLEX and the QMC
correspond to averages on $4\times 4$ ${\bf k}$-space
meshes. Actual QMC hole dopings averaged on $8\times 8$ ${\bf k}$-space
meshes are $9.7$, $14.0$, and $17.1\%$.
The real~(a) and the imaginary~(b) part.}
\label{fig_sig_flx}
\end{figure}

Fig.~\ref{fig_sig_flx} shows results for the FLEX calculations in
comparison with the QMC. FLEX approximation employed here uses
exchange of charge and spin fluctuations, i.e.,\ the
particle-hole-channel, but omits the particle-particle fluctuations
(for the Hubbard model there would only be the particle-particle
fluctuations of the singlet type). The results for the
real part are very similar to the screened-interaction expansion.
FLEX has a somewhat better qualitative behavior at high frequencies.
Low frequency behavior of the real part for
FLEX and screened-interaction expansions
are expected to be very similar because they are
both expected to obey the Luttinger theorem. The real part for
FLEX at the lowest frequency is doing somewhat worse than
the screened-interaction expansion, probably because ${\rm Im\:}\Sigma$
for FLEX is somewhat bigger, indicating that
FLEX results are more distant to their zero- or low-temperature values
[compare Figs.~\ref{fig_sig_scr}(b) and~\ref{fig_sig_flx}(b)].
${\rm Im\:}\Sigma$ is expected to vanish at zero
temperature and zero frequency also according to the
Luttinger theorem. FLEX results for the ${\rm Im\:}\Sigma$
show a correct qualitative behavior at high frequencies. Although
they also capture the doping dependence correctly, the magnitude
of the doping dependence is smaller by an order. In addition, they
don't do well quantitatively at any particular doping.

The screened-interaction-expansion
calculations performed in this article are
computationally very time consuming. The reason for this is the
requirement for the summation over many frequency and momentum
variables for the third and higher order diagrams. In order to
go to the fourth order, one would probably need Monte Carlo
summation techniques. Even at the third order, we were able to
perform our calculations on $4\times 4$ lattices only.
In Fig.~\ref{fig_sig_4X4} we make some comparisons of the
calculations on different lattice sizes. $4\times 4$ and $8\times 8$
QMC results are compared to the screened-interaction-expansion results.
FLEX results are shown as well.
The difference between the results for the two lattice sizes
seems to be very small. Estimated
QMC error bars for the ${\rm Im\:}\Sigma$ on the $8\times 8$ lattice
are also shown. These are the approximate upper bounds to
the statistical errors associated with the QMC calculations.

\section{EXACT QMC FERMI SURFACE}
\label{sec_fs}

In this section, we will discuss the exact QMC Fermi surfaces calculated.
Although there are several equivalent definitions
for the Fermi surface, the definition which locates the Fermi surface
as the zero frequency poles of the Green's function is most suitable
for finite temperatures. One finds the solution of
$\varepsilon({\bf k})+{\rm Re\:}\Sigma({\bf k},\omega=0)=\mu$ in order to
locate such poles.\cite{luttinger}
For a Fermi liquid, these poles correspond to
the quasiparticles at the Fermi surface. But, the definition
is general, and can also be used at finite temperatures. Note that,
the Luttinger theorem, which
relates the volume enclosed by the Fermi surface to the
total electron filling, is strictly valid only at zero temperature;
but, as long as the temperature is not unreasonably high, one
finds that the Luttinger theorem is satisfied in an approximate manner,
and the only change to the quasiparticle picture is some
temperature broadening due to the finite ${\rm Im\:}\Sigma({\bf k},\omega=0)$.
As mentioned in the previous section,
the quantity, ${\rm Im\:}\Sigma({\bf k},\omega=0)$, is indeed a
good indication of how close one is to $T=0$, which should
actually vanish at $T=0$ for a system obeying the Luttinger theorem.
In order to numerically calculate the Fermi surface on a finite $8\times8$
QMC lattice, we first find the solution of $\varepsilon({\bf k})+{\rm Re\:}
\Sigma({\bf k},\omega=0)=\mu$ by interpolating between the {\bf k} points.
Moreover, $\Sigma({\bf k},\omega=0)$, which is very close in value
to the one at the lowest
available frequency, $\Sigma({\bf k},\omega=i\pi T)$, is found by
linear extrapolation using the lowest-two frequencies.
Due to this numerical approximation procedures,
the systematic interpolation and extrapolation errors up to
a few percent are possible.

\begin{figure}[hbtp]
\begin{picture}(3.375,6.300)
\includegraphics{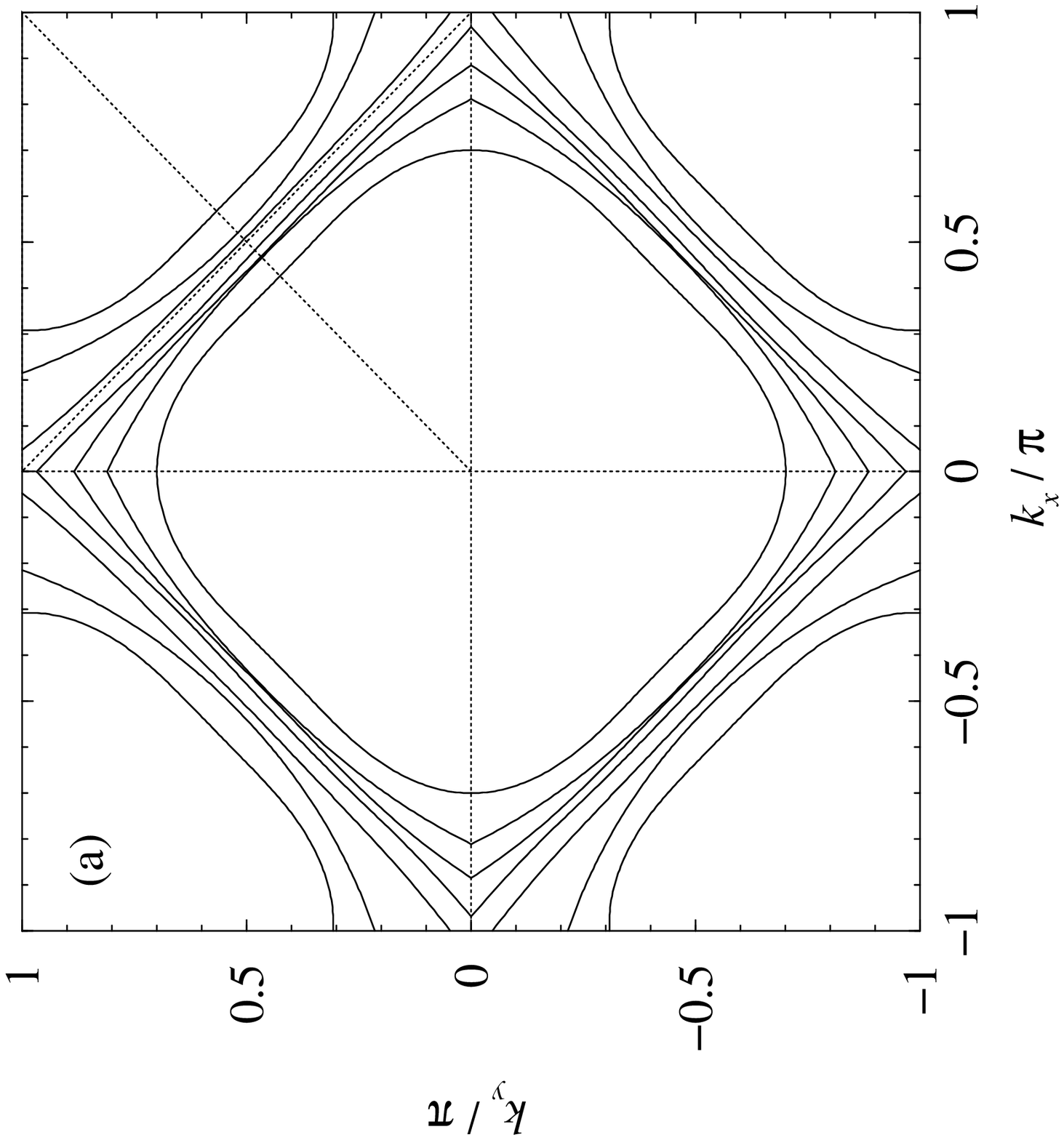}
\includegraphics{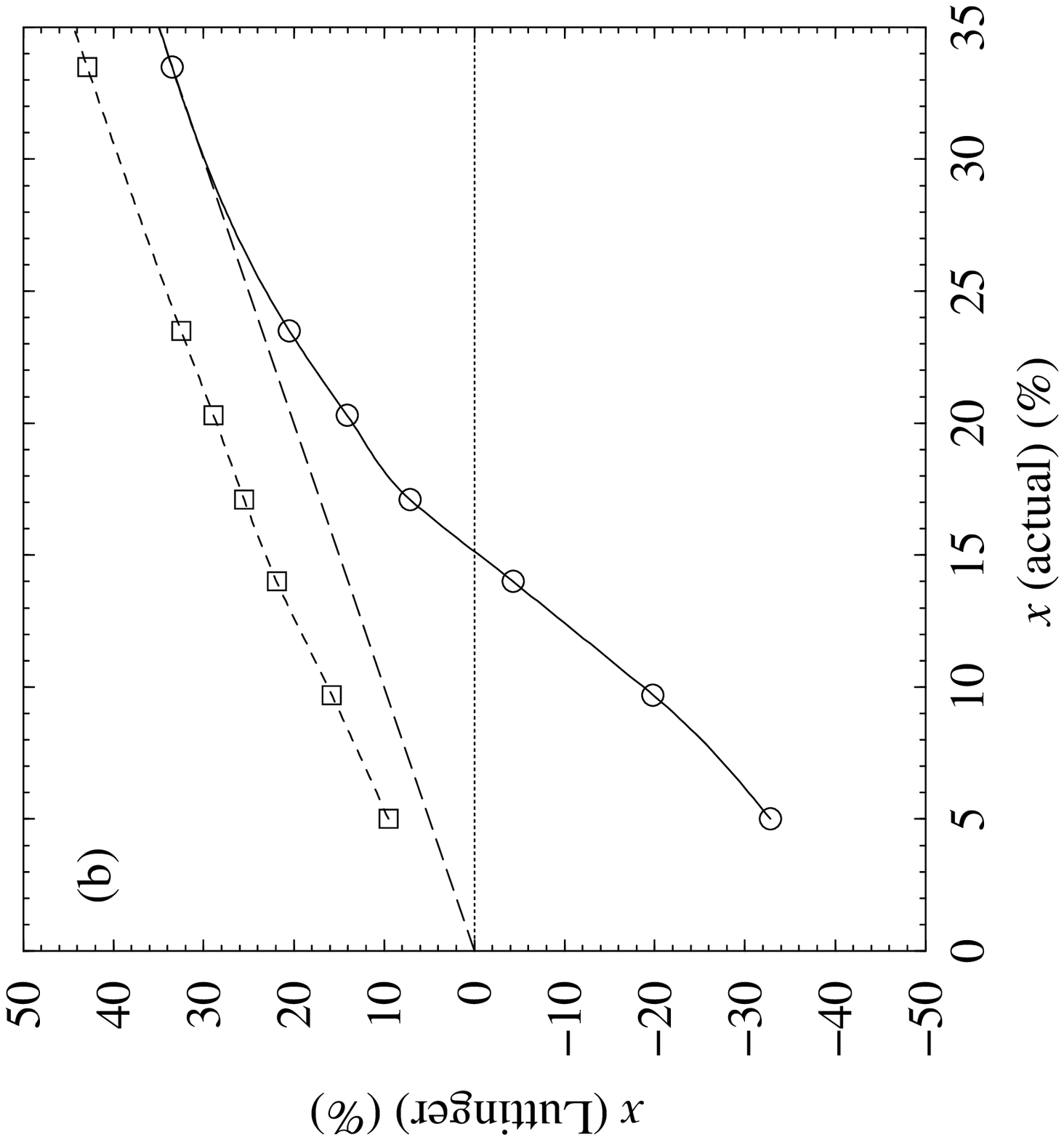}
\end{picture}
\caption{(a)~The exact QMC Fermi surfaces for various hole dopings.
Parameters are $U/t=8$ and $T/t=1/3$.
Actual hole dopings, $x\equiv 1-\langle n \rangle$,
in decreasing order are $x=33.5$
[smallest surface centered around ${\bf k}=(0,0)$],
$23.5$, $20.3$, $17.1$, $14.0$, $9.7$, and $5.0\%$
(the arcs centered around the Brillouin-zone corners).
The lattice size is $8\times8$.
(b)~The hole dopings, $x$, deduced from the areas enclosed by the Fermi
surfaces according to the Luttinger theorem, plotted against the
actual hole dopings, $x\equiv 1-\langle n \rangle$
(circles and the solid line).
Also shown are the FLEX results at the same temperature
(squares and the dashed line).
The long-dashed line without symbols corresponds to the Luttinger theorem.
The lattice size for the FLEX calculations is $16\times16$.}
\label{fig_fs}
\end{figure}

\begin{figure}[hbtp]
\begin{picture}(3.375,3.500)
\includegraphics{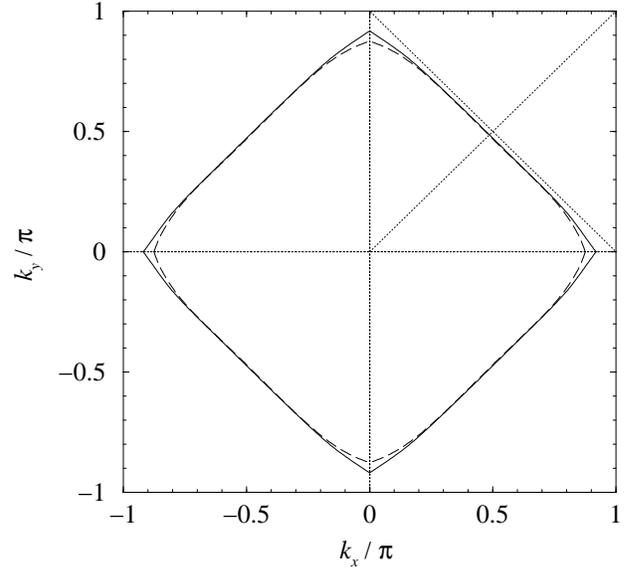}
\end{picture}
\caption{The FLEX Fermi surface at the temperature $T/t=1/3$
(long-dashed line) and at a lower temperature, $T/t=1/16$ (solid line).
This is the lowest available temperature for the lattice
size studied ($16\times 16$) at this doping. Parameters are $U/t=8$ and
$x\equiv 1-\langle n \rangle=5.0\%$. Hole dopings, $x$, deduced from the
areas according to the Luttinger theorem are
$7.6\%$ ($T/t=1/16$) and $9.5\%$ ($T/t=1/3$).
Note that the low-temperature FLEX Fermi surface is almost
the same as
the noninteracting (or Hartree-Fock) Fermi surface (not shown).}
\label{fig_fs_flx}
\end{figure}

Fig.~\ref{fig_fs}(a) shows the exact QMC Fermi surfaces
for actual
hole dopings, $x\equiv 1-\langle n \rangle$, ranging from $5.0$ to $33.5\%$.
These actual hole dopings in decreasing order are $x=33.5$
[smallest surface centered around ${\bf k}=(0,0)$],
$23.5$, $20.3$, $17.1$, $14.0$, $9.7$, and $5.0\%$
(the arcs centered around the Brillouin-zone corners).
We found the areas enclosed by the Fermi
surfaces by numerical integration, and calculated the hole dopings
implied by the Luttinger theorem corresponding to these areas. Note that,
due to the curve-fitting procedures employed for the discrete
$({\bf k},\omega)$ space, these areas are uncertain within
a few percent. The results in the same order as the actual hole
dopings are
$33.5$, $20.5$, $14.1$, $7.1$, $-4.3$, $-19.8$, and $-32.8\%$.
We plot these values against the actual doping in Fig.~\ref{fig_fs}(b).
Considering the fact that we are at a moderately high temperature,
$T=t/3$, and accounting for the curve-fitting errors,
the Luttinger theorem seems to be satisfied at least qualitatively
for hole dopings, $x\geq 17\%$, i.e.,\ the so-called overdoped
region in the context of high-temperature superconductivity.
On the other hand, in the underdoped region, QMC Fermi surfaces start
deviating from the Luttinger theorem in a systematic
qualitative matter. As one proceeds
towards half filling, Fermi surface starts shrinking towards the
${\bf k}=(\pi,\pi)$ point, looking like an electron-doped system.
This is completely inconsistent with the concepts of the band theory
and Fermi liquid. Fermi surface of the Hubbard model gradually gets smaller
and disappears when the hole doping gets smaller
in the underdoped region. Therefore, this seems to be a doping induced
transition between a
Fermi-liquid metal and a strongly-correlated insulator.
This extreme violation of the Luttinger theorem may imply that there
is no asymptotically convergent diagrammatic expansion for the
Hubbard model in the underdoped region. But, it looks like,
as the hole doping increases past the optimal doping ($\sim 15\%$),
the system looks more like a conventional Fermi liquid, and diagrammatic
expansions seem more likely to be possible.
In Fig.~\ref{fig_fs}(b), we plot the hole dopings, $x$, deduced from
the area enclosed by the Fermi surface according to the Luttinger
theorem against the actual hole dopings. Systematic deviation
towards half filling is clearly seen. Note that the line is only
a guide to the eye; it is quite possible that the area of the
QMC Fermi surface at dopings
very close to half filling gets vanishingly small.
We don't have data for such dopings at the point, so, further
investigation is necessary to determine the existence of
such ``hole pockets.''\cite{holepockets}
In order to rule out temperature
effects one might also study lower temperatures, which could
be accessible for such dopings very close to half filling.
This figure also shows the results for FLEX calculations
at the same temperature. Since FLEX is expected to satisfy
the Luttinger theorem at low temperatures, the deviations
in this case are known to be mainly because of temperature effects.
But, these deviations are much more reasonable, and the FLEX
results show approximately correct slope for all dopings.
Doping deduced from the Luttinger theorem for FLEX calculations
also have the correct sign for all dopings. We show the FLEX
Fermi surfaces at $5.0\%$ doping for the temperature studied
above ($T=t/3$) along with a low-temperature Fermi surface
in Fig.~\ref{fig_fs_flx}. The low- and intermediate-temperature Fermi
surfaces for FLEX look very similar, and the low-temperature
case is almost identical to the noninteracting case (for the Hubbard
model, which is also the same as the Hartree-Fock case). The comprehensive
results we
obtained for the Fermi surface are also qualitatively consistent
with the other recent QMC study of the Hubbard\cite{qmc} and high-temperature
expansion study of $t$-$J$\cite{singh} models.

\section{SUMMARY AND CONCLUSIONS}
\label{sec_sum}

In this article, we studied a new approximation for the one-particle
self energy of the large-$U$ Hubbard-like models. An expansion was
written out in terms of a screened interaction, which is nonlocal
and retarded, although it acts between local and instantaneous
charge densities. The screening is achieved by the charge fluctuations.
The expansion is an exact perturbative expansion, although this
doesn't necessarily mean that it is an exact solution because
of the possibility of nonperturbative nature of the
large-$U$ systems. We carried out the expansion up to the third order.
The screened interaction was determined by QMC calculations exactly,
and then directly put into the expansion calculations.
The local component of the
exact screened interaction, which screens the bare Hubbard $U$,
was found to be vanishing
at certain small frequencies at hole dopings near $15\%$.
This is a very encouraging behavior for a weak-coupling expansion,
which indicates that such expansions might be possible
at low-enough frequencies where the screened-interaction is very small.
However, the analysis of the three-point-vertex function, which, together
with the screened interaction,
determines the one-particle self energy, showed that it diverges
at the same frequency-momentum points where the screened interaction
vanishes. The expansion for the self energy had only limited success
against the exact QMC results. Although the expansion gave excellent
results for the low-energy part of
${\rm Im\:}\Sigma$ near $15\%$ hole doping, where
the screened interaction vanishes, it had the wrong doping dependence
around this same value. Moreover, the real part of the QMC results
were not consistent with a diagrammatic expansion which should obey
the Luttinger theorem, although, we can't rule out the possibility
of temperature effects. If it were because of temperature
effects, this would also mean that our approximation doesn't work
for ${\rm Re\:}\Sigma$ at all, since with the increasing order,
results for the expansion moved in the wrong direction. In contrast,
${\rm Im\:}\Sigma$ showed a converging behavior near $15\%$ doping,
as mentioned above. We also calculated FLEX self energies,
which had the correct qualitative doping dependence and high-frequency
behavior, but did actually worse quantitatively near $15\%$ doping.

In order to investigate the QMC ${\rm Re\:}\Sigma$ results,
we numerically calculated the exact QMC Fermi surfaces. The results
showed extreme violation of the Luttinger theorem near half filling,
which might be because of a doping-induced metal-insulator transition
arising from strong-coupling effects. The Fermi surface of the Hubbard
model seemed to be reducing to hole pockets centered around ${\bf k}=(\pi,\pi)$
at small hole dopings near half filling. Further QMC investigation
at smaller hole dopings and lower temperatures might bring more evidence
into this.

As far as the high-temperature superconductors are concerned,
exact QMC results might be indicating that the underdoped
region of these materials is a transition between a strong-coupling
Mott-Hubbard insulator and a Fermi metal. The disappearance
of the Fermi surface with
the decreasing hole doping can also be viewed
as an opening of a pseudogap.\cite{pseudogap}
For the full investigation
of this region at low temperatures,
other nonperturbative treatments might be necessary.
However, the partial success of the screened-interaction expansion
developed here shows that it might be useful near optimal doping, as well as
other diagrammatic expansions in the optimally-doped and overdoped
regions.

\acknowledgments

G.~E. acknowledges useful discussions with Richard~T. Scalettar
and Andrew~K. McMahan.
This work was partially supported by National Science Foundation under
the Grants No.\ DMR-92-15123 and No.\ DMR-99-70291.
Work at University of California, Davis
was supported in part by an Accelerated Strategic Computing Initiative Grant
and by Materials Research Institute of Lawrence Livermore National Laboratory.
Computing support from University Computing and Networking Services
at University of Georgia is gratefully acknowledged.


%
%

%
%


\begin{references}

\bibitem[*]{address_GE}
Present address: Department of Physics, University of California,
Davis, California 95616-8677; and Lawrence Livermore National Laboratory,
University of California, Livermore, California 94550-9234

\bibitem[\dag]{email_GE}
E-mail address: esirgen@ucdavis.edu

\bibitem[\ddag]{address_HGE}
Present address: Institut f\"{u}r Theoretische Physik, Technische
Universit\"{a}t Graz, 8010 Graz, Austria

\bibitem{hubbard}
E. Dagotto, Rev.\ Mod.\ Phys.\ {\bf 66}, 763 (1994).

\bibitem{mis}
M. Imada, A. Fujimori, and Y. Tokura, Rev.\ Mod.\ Phys.\ {\bf 70}, 1039 (1998).

\bibitem{ed}
W. Fettes, I. Morgenstern, and T. Husslein, Computer Physics Communications
{\bf 106}, 1 (1997); H.~Q. Lin, J.~E. Hirsch, and D.~J. Scalapino,
Phys.\ Rev.\ B {\bf 37}, 7359 (1988).

\bibitem{qmc}
C.~Gr\"{o}ber, R. Eder, and W. Hanke, Phys.\ Rev.\ B {\bf 62}, 4336 (2000).

\bibitem{flex1}
G. Esirgen, H.-B. Sch\"{u}ttler, and N.~E. Bickers,
Phys.\ Rev.\ Lett.\ {\bf 82}, 1217 (1999); and references therein.

\bibitem{HSSJ}
M.~S. Hybertsen, E.~B. Stechel, M. Schl\"{u}ter, and D.~R. Jennison,
\prb {\bf 41}, 11068 (1990);
C.-X. Chen and H.-B. Sch\"uttler, {\em ibid.}\ {\bf 43}, 3771 (1991);
S.~B. Bacci, E.~R. Gagliano, R.~M. Martin, and J.~F. Annett,
{\em ibid.}\ {\bf 44}, 7504 (1991).

\bibitem{ScFe} 
H.-B. Sch\"uttler and A.~J. Fedro, \prb {\bf 45}, 7588 (1992).

\bibitem{sgeh}
H.-B. Sch\"{u}ttler, C. Gr\"{o}ber, H.~G. Evertz, and W. Hanke,
cond-mat/9805133; 0104300 (to be published).

\bibitem{luttinger}
J.~M. Luttinger, Phys.\ Rev.\ {\bf 119}, 1153 (1960).

\bibitem{holepockets}
A.~P. Kampf and J.~R. Schrieffer, Phys.\ Rev.\ B {\bf 42}, 7967 (1990);
S.~A. Trugman, Phys.\ Rev.\ Lett.\ {\bf 65}, 500 (1990);
D. Duffy and A. Moreo, Phys.\ Rev.\ B {\bf 51}, 11882 (1995).

\bibitem{singh}
W.~O. Putikka, M.~U. Luchini, R.~R.~P. Singh, Phys.\ Rev.\ Lett.\ {\bf 81},
2966 (1998).

\bibitem{pseudogap}
For experimental evidence of the pseudogap, see, e.g., the following
and references therein: Z.-X. Shen and J.~R. Schrieffer,
Phys.\ Rev.\ Lett.\ {\bf 78}, 1771 (1997) [ARPES];
R. Nemetschek, M. Opel, C. Hoffmann, P.~F. M\"{u}ller, R. Hackl,
H. Berger, L. Forr\'{o}, A. Erb, and E. Walker,
Phys.\ Rev.\ Lett.\ {\bf 78}, 4837 (1997) [Raman];
G.~V.~M. Williams, J.~L. Tallon, E.~M. Haines, R. Michalak, and R. Dupree,
Phys.\ Rev.\ Lett.\ {\bf 78}, 721 (1997) [NMR].

\end{references}
\end{document}